\documentstyle[preprint,aps,prd]{revtex}

\tighten
\begin{document}

\title{Finding Apparent Horizons in Dynamic 3D Numerical Spacetimes}
 
\author{
Peter Anninos$^{1}$, Karen Camarda$^{1,4,6}$, Joseph Libson$^{1,4}$, Joan
Mass\'o$^{1,2,6}$, Edward Seidel$^{1,4,6,7}$, Wai-Mo Suen$^{3,5}$}

\address{
$^1$ National Center for Supercomputing Applications \\
605 E. Springfield Ave., Champaign, Illinois 61820}

\address{
$^2$ Departament de F\'\i sica, Universitat de les Illes
Balears, \\ E-07071 Palma de Mallorca, Spain}

\address{
$^3$ McDonnell Center for the Space Sciences, Department of Physics
Washington University, St. Louis, Missouri, 63130}

\address{
$^4$ Department of Physics, University of Illinois at
Urbana-Champaign, Illinois 61801}

\address{
$^5$ Physics Department, Chinese University of Hong Kong, Shatin, Hong
Kong}

\address{
$^6$ Max-Planck-Institut f{\"u}r Gravitationsphysik, Schlaatzweg 1, 14473 Potsdam, Germany}

\address{
$^7$ Department of Astronomy, University of Illinois
Urbana, Illinois 61801}

\date{\today}
\maketitle

\begin{abstract}
We have developed a general method for finding apparent horizons in 3D
numerical relativity.  Instead of solving for
the partial differential equation describing the location of the
apparent horizons, we expand the closed 2D surfaces
in terms of symmetric trace--free tensors and solve for the
expansion coefficients using a minimization procedure.
Our method is applied to a number of different spacetimes, including
numerically constructed spacetimes containing highly distorted
axisymmetric black holes in spherical coordinates, and 3D
rotating, and colliding black holes in Cartesian coordinates.
\end{abstract}


\section{Introduction}
\label{sec:intro} 

Black holes are among the most fascinating predictions in the theory
of General Relativity.  The black holes most likely to be observed by 
future gravitational wave
observatories (LIGO and VIRGO~\cite{LIGO3}) are those
in highly dynamical evolutions, such as
two colliding black holes. Moreover, events which are
important for observations (i.e., events that occur more frequently
and emit stronger radiation) are not expected to have a high
degree of symmetry; for example, the inspiraling coalescence is a more 
probable scenario than the
axisymmetric head--on collision of two black holes.  The most powerful
tool in studying such highly dynamical and intrinsically non-linear
events is numerical treatment.

The essential characteristics of a black hole are its
horizons, in particular, the apparent horizon (AH) and the event
horizon (EH).  One needs to determine the location and
the structure of the EH's in numerical studies to
understand the properties of black holes, and indeed even 
to assert the existence of the holes.
Algorithms for doing this have recently been
developed~\cite{Hughes94a,Libson94a}.  
In contrast, the problem of determining
the location of the AH in a general numerically constructed 3D
spacetime has not yet been solved satisfactorily.  The present paper
represents a step in this direction.

The apparent horizon is defined to be the outer-most marginally
trapped surface~\cite{Hawking73a}, a surface for which the divergence
of the out-going null normal is zero (c.f., eq. (1) below).  The
surface is defined locally in time, in contrast to the EH,
which can only be identified after the numerical simulation is
complete.  The AH, as a characteristic of
black holes, can be used {\it during} the numerical construction of the
spacetime.  As discussed in a number of publications, apparent
horizons are useful not only for studying the dynamics of black hole
spacetimes\cite{Anninos93a}, but also for use as an inner boundary
in numerical evolutions of black 
holes\cite{Thornburg93,Seidel92a,Anninos94e}.
The so-called apparent horizon boundary condition (AHBC)
is currently being developed by many groups as a promising
method for use in computing the long term evolution of 3D black hole
systems.  With AHBC one would like to be able to track
the AH throughout the numerical evolution.
For these reasons, it is important to develop efficient
methods for locating apparent horizons in numerically constructed
spacetimes.

There are many well developed methods for determining the
location of AH's in lower dimensional 
spacetimes~\cite{Cadez74,Bishop82,Cook90a,Tod91,Bernstein93a},
e.g., in axisymmetry.
The partial differential equation (PDE) defining a
marginally trapped surface (eq. (1) below) reduces to an ordinary
differential equation (ODE) in the axisymmetric case, and the
symmetry conditions also provide boundary conditions for starting the
integration of the ODE.  This simplifies tremendously the problem, and
enables the construction of efficient methods for finding the AH.
However, as these methods rely strongly on the symmetry assumptions,
they are not generalizable to 3D; going from 2D to 3D does not amount to
simply adding one more spatial dimension.
For the general 3D case, there is no symmetry and the AH surface to be
determined is a {\it closed } surface (hence no boundary conditions
for starting the integration) described by a non-linear elliptical
PDE.  At present there are no efficient
algorithms for solving such a partial differential problem in general.

We are aware of three independent efforts in determining the AH in the
general 3D case.  The first method is based on an expansion of the AH
surface in terms of spherical harmonics, with the expansion
coefficients determined by an integral equation.  The equation is then
solved iteratively~\cite{Nakamura84,Kemball91a}.  The second method
attempts to solve directly the elliptic PDE~\cite{Thornburg95,Huq96}.
The third method~\cite{Libson95a} is based on expanding the closed
surface in terms of orthogonal functions, in particular symmetric
trace--free (STF) tensors, and using a minimization procedure to
determine the expansion coefficients. Variations on this idea were
explored by Brill and Lindquist~\cite{Brill63} and by
Eppley~\cite{Eppley77}.  The essential difference of this method with
the first method is in the way the expansion coefficients are
determined, and the use of STF tensor expansions for 3D Cartesian
codes .  The major advantage of this method is that the numerical
solution of the minimization problem is much better understood than
the corresponding PDE problem.  In Ref.~\cite{Libson95a}, we
demonstrated that our method can be efficiently applied to testbeds
made up of analytically given data sets representing time--symmetric
slices of spacetime.  In Ref.~\cite{Baumgarte96a}, the convergence of
the symmetric trace--free tensor expansion for similar testbeds is
studied in detail.  In this paper, we follow up on our earlier work
and give a more detailed discussion of this method. We also push the
application of it in two directions: (1) Application of the method to
arbitrary data sets which are not time--symmetric (non--zero extrinsic
curvature), and (2) application of the method to data obtained in
actual numerical evolutions of dynamical spacetimes.  The spacetimes
we studied include Schwarzschild and Kerr black holes, black hole plus
Brill wave, and Misner two black hole spacetimes in full 3D.  We
demonstrate that our method is in principle applicable to general 3+1
spacetimes.  Another method presently
under development also uses a series expansion as in the present method,
but it evolves the trial surface to the AH in an algorithm that
combines elements of the AH finder of ~\cite{Nakamura84} and of a
curvature flow method ~\cite{Gundlach97a}.

In \S \ref{sec:implement} 
we discuss the formulation of our method, and the numerical
algorithm in detail.  Section \ref{sec:testbeds}
gives the results of various testbed calculations,
with \S \ref{subsec:initial}
concentrating on initial data sets, and \S \ref{subsec:evolved} on 
spacetime evolutions.  Section \ref{sec:conclusion} concludes with a
discussion of where we stand in the
construction of a general method for finding 
apparent horizons in 3D numerical relativity.

\section{Formulation of the AH Finder}
\label{sec:implement}

\subsection{Basic Equations}
\label{subsec:equations}

Defining $s^\mu$ to be the outward--pointing spacelike unit normal of
a two--sphere $\cal S$ embedded in a constant time slice $\Sigma$
with unit normal $n^\mu$, we can construct the outgoing null normal to
any point on $\cal S$ as $k^\mu = n^\mu + s^\mu$. The surface $\cal S$
is called a {\it marginally trapped surface} (MTS) if the divergence of the
outgoing null vectors vanishes $\nabla_\mu k^\mu = 0$, or equivalently
\cite{York89}
\begin{equation}
\Theta = D_i s^i + K_{ij}s^i s^j - K = 0 ,
\label{expansion}
\end{equation}
where $\Theta$ is the expansion of the outgoing rays,
$D_i$ the covariant derivative with respect to the 3--metric $\gamma_{ij}$,
$K_{ij}$ the extrinsic curvature of $\Sigma$, and $K$ the trace
of $K_{ij}$.
The AH is defined as the outer--most trapped surface.

First suppose we are searching for the AH of a single black hole in
spherical coordinates.
As the AH is topologically a 2--sphere \cite{Gibbons72}, 
its position can be represented as
\begin{equation}
F(\theta, \phi, r) = r^2
                     - f(\theta, \phi) = 0 .
\label{capF}
\end{equation}
The unit normal $s^i$ 
\begin{equation}
s^i = \gamma^{ij} \partial_j F 
      \left(\gamma^{k\ell} \partial_k F \partial_{\ell} F \right)^{-1/2} ,
\label{s}
\end{equation}
can then be expressed in terms of the function  $f( \theta , \phi)$. 
Substituting (\ref{capF}) and (\ref{s}) into equation (\ref{expansion}), 
one gets an elliptic
equation for $f(\theta, \phi)$.  Instead of solving this elliptic
equation directly, we proceed by expanding $f$ 
in terms of the usual spherical harmonics $Y^{\ell m}$:
\begin{equation}
f(\theta, \phi) = \sum_{\ell=0}^L \sum_{m=-\ell}^{\ell} 
                   F^{\ell m}Y^{\ell m}(\theta, \phi) .
\label{exf}
\end{equation}
Equation (\ref{expansion}) then gives
$\Theta$ in terms of the expansion coefficients $ F^{\ell m}$.
The AH can be determined by finding the set of 
coefficients {$ F^{\ell m}$} which make $\Theta =0$.

For 3D codes in Cartesian coordinates, instead of the spherical harmonic
expansion (\ref{exf}), we choose to
expand the trial surfaces
in terms of symmetric trace--free (STF) tensors:
\begin{equation}
F(x,y,z) =  \sum_{i=1}^3 \left(x^i - x_0^i\right)^2 
- \sum_{\ell=0}^L {{\cal F}_{K_\ell}} N_{K_\ell}  = 0 ,
\label{stf1}
\end{equation}
where $x^i$ are Cartesian coordinates and $x_0^i$ the Cartesian points
interior to the $F=0$ surface representing the horizon center.
In expanding the function $f$ in equation (\ref{stf1}), we
have adopted the notation from reference \cite{Thorne80}.  The STF
tensors ${\cal F}_{K_\ell}$ are coordinate independent coefficients, the
subscript $K_\ell$ is abbreviated notation for the vector product
$N_{K_\ell} \equiv n_{k_1} n_{k_2} . . . n_{k_\ell}$, and the $n_i$
are unit directional vectors
\begin{equation}
n_i = \frac{x^i - x_0^i}{\sqrt{\sum_{j=1}^3 (x^j-x_0^j)^2}} .
\end{equation}

To determine the set of  ${\cal F}_{K_\ell}$ which makes (5) the AH, we
use a minimization procedure.
In general, the AH surface is the outer-most surface represented by
the set of ${\cal F}_{K_\ell}$ and $x_0^i$ which satisfy
\begin{equation}
\sum_\sigma W_\sigma \Theta_\sigma^2({\cal F}_{K_\ell}, x _ 0 ^ i) = 0 .
\label{the}
\end{equation}
$\Theta^2$ is semi--positive definite.
$W_\sigma$ is
a positive definite weighting function which can
be chosen to improve the accuracy of the minimization procedure depending
on the construction of the trial surface.  
In this paper we do not investigate the use of this parameter 
and set $W_\sigma = 1$.
With this we have converted the elliptic condition (1) to a standard
minimization problem.  There are 
efficient minimization algorithms to search the space of (${\cal F}_{K_\ell}, 
x_0^i$) coefficients for the surface closest to the apparent horizon 
among all test surfaces so parameterized.  The strength of our method
lies in that the minimization problem is much better understood than
the numerical solution of the corresponding differential equation (1).

An obvious potential difficulty of this method is that there is no
guarantee that the summation (\ref{stf1}) converges, or 
converges rapidly.  For black holes not in highly dynamical situations,
we do expect the AH surface to be smooth and one needs only the first
few lower rank tensors to find the surface accurately (as demonstrated
in section \ref{sec:testbeds}).  This is usually the case 
through most of the simulations, and our method will be
more efficient at these times.
However, we also expect that, for the cases we
would like to simulate, there are often periods of time when the
black hole (or holes) goes through highly dynamical evolutions, e.g.,
around the moment of the coalescence of two black holes.  In such
cases, one needs higher order STF expansions.  Although STF tensor
expressions are available in the literature, we find it convenient to
include in our code a routine for the automatic generation of STF
tensors to an arbitrarily high order. The
procedure takes advantage of the fact that we can associate the $n_i$
with essentially $x-x_0$, $y-y_0$ or $z-z_0$ in three dimensions.  We
can therefore construct all symmetric and independent permutations of
\begin{equation}
N_{K_\ell} \equiv A _ l (x-x_0)^\alpha (y-y_0)^\beta (z-z_0)^\gamma ,
\end{equation}
subject to the constraint $\alpha+\beta+\gamma=\ell$, where $\ell$ is
the rank of the corresponding symmetric tensor, or equivalently the
order of the multipole expansion, and $A _ l$ is the normalization
factor making the RHS dimensionless.  
There are $(\ell+1)(\ell+2)/2$ such independent combinations.
The combinations constructed in this fashion can be supplemented with
the $\ell (\ell-1)/2$ independent conditions imposed on the symmetric
permutations to make the rank $\ell$ tensor trace--free by contracting
on any two indices. The function $f$ can be expanded as
\begin{equation}
f(x,y,z) = \sum_{\ell=0}^L \sum_k^{\alpha+\beta+\gamma=\ell}
                  C_k < (x-x_0)^{\alpha_k} (y-y_0)^{\beta_k}
                      (z-z_0)^{\gamma_k} >,
\label{multipole}
\end{equation}
where $C_k$ are coordinate independent coefficients, and $< >$ denotes
STF combinations \cite{Thorne80}. Partial derivatives
$\partial_i F$ and $\partial_i \partial_j F$ needed to evaluate 
$\Theta$ on the trial surfaces are then easily computed
from (\ref{capF}) and (\ref{multipole}).
The simplicity of the form (\ref{multipole}) also allows one to easily
construct the multipole expansion to take advantage of any
symmetries present in the problem. For instance, our current
implementation, by setting a flag, can invoke either the even or odd
multipoles independently of the other, enforce axisymmetry, 
fix the surface centers $x_0^i$,
or allow the most general parametric expansion.

\subsection{Numerical Algorithms}
\label{subsec:numerical}

The numerical problem is to find a set of parameters ($C_k$, $x_0^i$)
that minimizes the LHS of (\ref{the}).  Minimization techniques (such
as conjugate gradient or quasi-Newton methods) that evaluate both the
function and all its various partial derivatives are often preferred,
as a means to increase the convergence rate, to those that do not
require derivative information.  However, because of the complexity of
equation (\ref{expansion}), in this first generation 3D AH finder we
have chosen to implement a multi--dimensional method that does not
require knowledge of derivatives, namely a direction set or Powell's
method \cite{Press86}.  The method is based on successive line
minimizations, whereby the function $\sum \Theta^2$ is successively
minimized along different vector directions using the one--dimensional
Brent's method with parabolic interpolation.  We find Powell's method
to be generally robust, with a good convergence rate and computational
speed, for surface functions parameterized by fewer than about fifteen
or so parameters (see section \ref{sec:testbeds}).  Although solutions
can still be found for parametrizations of higher order, the
computational cost becomes overly excessive, especially when compared
to the evolution cycle time.

A two--grid system is utilized to evaluate the expansion function $\Theta$.
The first grid (which we refer to as grid $A$) 
is the Cartesian--based  computational grid on which the
Einstein evolution and constraint equations are solved, and the 
metric and extrinsic curvature components are defined. A second grid $B$
is used to evaluate the surface function $F(\theta, \phi, r)$. This
second grid need not be structured in the same way as the first.

Assuming a guess position for the horizon center $x_0^i = (x_0, y_0, z_0)$
on the structured Cartesian grid $A$,
and some prescription for distributing points throughout this
mesh to construct nodes for the second grid $B$, we can evaluate
the expansion $\Theta$ on the nodes of grid $B$
once $f(\theta, \phi)$ is determined. 
Here we choose to distribute the nodes uniformly over a sphere, 
centered on $x_0^i$:
the nodes are evenly spaced in both
the polar and azimuthal angles to cover the full sphere 
$0\le \theta \le \pi$ and $0\le \phi \le 2\pi$ (or 
a single octant for the axi-- and equatorial symmetric spacetimes).
However, this procedure can easily be generalized
to, for example, weight the node distribution according to the 
coordinate surface curvature, as might be desirable for
highly distorted horizons. Along the radial direction, the nodes of grid $B$
are placed uniformly with an inter--node spacing $\delta r$ 
typically equal to the
cell resolution of grid $A$ on which the Einstein equations are solved.
We have also implemented a procedure to constrain the range of
radii over which the solver searches for the AH, as might be
desirable in cases where multiple trapped surfaces exist in the data set.
In these cases, the radial grid spacing is set by
$\delta r = (r_{max} - r_{min})/N_r$, where $N_r$ is the number
of radial nodes, and $r_{min}$ and $r_{max}$ are the lower and upper
bounds of allowable radii.
Representing the number of nodes on grid $B$ as 
$N_\theta \times N_\phi \times N_r$, we set 
$N_\theta = N_\phi = 5$ along the angular directions in a single octant,
and $N_r = N$ along the radial direction,
where $N$ is the minimum number of cells among the three 
orthogonal axes in grid $A$.

Once the spherical grid $B$ is constructed and centered on $x_0^i$,
the function $F$ in equation (\ref{capF}) 
is evaluated on the nodes of grid $B$ for a fixed set
of coefficients ($C_k, x_0^i$) that the Powell routines compute.
Along each radial line, a search is made for $F=0$, by scanning from
large to small radii, until the condition
$F(\theta,\phi,r) \times F(\theta,\phi,r+\delta r) < 0$ is met, and then
linearly interpolating between adjacent neighbors to find the approximate
Cartesian coordinates of the surface corresponding to $F=0$.
The extrinsic curvature and the metric functions 
and their derivatives are then evaluated at these
positions by interpolating (either linearly or quadratically) from the
computational grid $A$ on which they are defined
and used in equation (\ref{expansion}) to evaluate $\Theta^2$
on the surface. The process of constructing a spherical grid $B$
centered on $x_0^i$, evaluating $F$ on grid $B$, searching for the
surface coordinates for which $F=0$, interpolating the geometric
data to the surface, and evaluating the expansion (\ref{expansion})
on the surface is repeated through Powell's procedure until a
minimum of $\sum\Theta^2$ in the parameter space ($C_k, x_0^i$) is realized.

To reduce the computational time spent in finding apparent horizons
and to make the finder usable in real time evolutions of black hole
spacetimes, it is important to implement the solver in a
parallel fashion.  Fortunately, the calculations performed at each
node on the trial 2D surfaces are independent of the other node
calculations. A natural parallel implementation, which we have
adopted, thus distributes the different surface calculations to
different computational processors, achieving a speedup of $N_\theta
\times N_\phi$ compared to a purely scalar code. In addition, a large
percentage of the cpu time is spent interpolating the field variables
from grid $A$ onto the 2D surface. To speedup this bottleneck
procedure, we have written and implemented generalized algorithms
designed to interpolate (both linearly and quadratically) 3D data onto
different nodes within a 2D surface in parallel.  The
computationally intensive elements of the horizon finder have 
been optimized for both the Cray C90 and the massively parallel
Thinking Machines CM5 computers.

\section{Code Tests}
\label{sec:testbeds}

In order to test the basic solver described above, we have developed
both 2D and 3D AH finders based on this minimization method.

For the 2D finder, the AH determined can be directly compared to those
obtained with the standard integration method.
As testbeds, we used data obtained from a code developed by Bernstein
{\em et al.}\cite{Bernstein93b}.  This code evolves a black hole
distorted by an axisymmetric distribution of gravitational waves
(Brill waves)\cite{Bernstein94a}.  The black holes can be highly
distorted by the incoming waves, leading to AH's with extremely prolate
or oblate geometric shapes.  In some cases the ratio of polar to
equatorial circumference can exceed $10^2$.  When these systems are
evolved, the horizons undergo dynamic oscillations, eventually
settling down to a Schwarzschild black hole at late
times \cite{Anninos93a}.  For such dynamic spacetimes, we compared the
results obtained with our new AH finder algorithm with those 
from the AH finder constructed using the standard ODE
method \cite{Bernstein93a,Anninos93a}.  
For test surfaces with 16 coefficients,
and using spherical harmonics $ Y_{lm}$ as basis functions,
we find that both methods produce the same results to
within the given accuracy of the PDE solvers.

As this paper is on a 3D implementation of these ideas for finding AH's,
in the following we concentrate on results for the 3D case only.  
We have written a 
Fortran routine that implements the above method for a 
general 3D spacetime in Cartesian coordinates,
and tested it on various spacetimes of interest. 
We discuss results from
this code applied to various initial data sets
containing one or more black holes (\S \ref{subsec:initial}), 
and to evolutions of 
some of these black hole spacetimes carried out with our 3D ``G'' code 
\cite{Anninos94c} (\S \ref{subsec:evolved}). For all the 3D tests, we 
use the symmetric trace--free (STF) tensors as basis functions defined
on a unit sphere.

\subsection{Finding Horizons in Initial Data Sets}
\label{subsec:initial}

\subsubsection{Schwarzschild Black Hole}
\label{sec:schwarzinit}

The simplest, most basic, test for any apparent horizon finder is the
static Schwarzschild initial data. The 3-metric can be written in
Cartesian coordinates as
\begin{equation}
d\ell^2= \left(1+\frac{M}{2r}\right)^4\left(dx^2+dy^2+dz^2\right),
\end{equation}
where $M$ is the 
mass of the black hole. The apparent horizon in this case is
spherically symmetric and located at $r=M/2$. Although only
the $L=0$ term should contribute to the surface that defines the
apparent horizon, we tested the solver with a more general multipole
expansion with the $m\ne 0$ terms up to and including $L=6$, a total of
28 coefficients.

The computed surface is plotted in Fig.~\ref{fig:sch_init} for four
separate cases with various black hole masses.  The grid resolution
used in each case was $\Delta x = \Delta y = \Delta z = 0.075M$, using
$25^3$ cells.  As expected, the surface is mostly defined by the
$\ell=0$ contribution: The other higher order terms are small in
comparison, roughly a factor of $10^{-8}$ smaller. We find typical
errors in the horizon radius of order 0.04\%, and the numerical
surfaces are indistinguishable from the analytic solution in
Fig.~\ref{fig:sch_init}.  In each of the cases shown here, the finder
converged to the correct surface in approximately 30 iterations.
However, the number of iterations decreases significantly if fewer
parameters are varied. For example, only 3 iterations are needed to
converge if only the monopole term is varied, reducing the
computational time by a factor of 60.  On average, the cpu time scales
approximately as $N_p$, where $N_p$ is the number of
parameters. Hence, the method becomes rather cumbersome for highly
distorted horizons which can only be described with a high order
multipole expansion. We discuss this important issue further in the
following more elaborate tests.

\subsubsection{Misner Data}
\label{sec:misnerinit}

The Misner initial data set represents two equal mass black holes
initially at rest, and is defined by the 3--metric
\begin{equation}
d\ell^2= \psi^4\left(dx^2+dy^2+dz^2\right),
\end{equation}
where
\begin{equation}
\psi=1+\sum_{n=1}^{\infty}\frac{1}{\sinh(n\mu)}
       \left(\frac{1}{{}^{+}r_n}+\frac{1}{{}^{-}r_n}
       \right),
\end{equation}
and
\begin{equation}
{}^{\pm} r_n=\sqrt{x^2+y^2+\left[z\pm \coth(n\mu)\right]^2}.
\end{equation}
The parameter $\mu$ specifies the proper separation between the
two holes and the total ADM mass of the spacetime.
Apparent horizons in these data sets may consist of either a
single surface surrounding both black holes if they are
sufficiently close to one another ($\mu < 1.36$) \cite{Smarr76}, 
or two separate horizons located at the throats of the holes.

In the cases where the two holes form a single encompassing
horizon, the surface can be distorted significantly.
To find the distorted horizons accurately, then, 
we need to keep higher order multipole terms, but {\em a priori} it is
not clear how many terms will be required.  In
Fig.~\ref{fig:mis_init}, we show the results of systematically
increasing the number of axisymmetric expansion terms for the case
$\mu=1.2$.  In each calculation we use a $64^3$ grid with
$\Delta x = 0.1$ and run the code on the 128 node partition of the
CM5. We also show the result obtained with our 2D, axisymmetric
code described in Ref.\cite{Anninos94b}, which implements an
independent ODE integration method~\cite{Anninos93a}.  
It is obvious that a high order expansion, up to
$L=6$, is required to accurately describe this surface which 
has a major to minor axis ratio of 1.5.
We expect that even higher multipole expansions would be required
for more distorted surfaces.
In Table \ref{tab:misner}, we show the number
of iterations, cpu time and $\sum\Theta^2$ as a measure
of convergence for each of the expansion orders.
We note that timings reported here and throughout this paper
refer to the Thinking Machines CM5.

For $\mu>1.36$, there are two separate and spherical trapped surfaces
on the initial slice, centered off the origin
at $z=\pm \coth\mu$. In the cases we have
tested ($\mu=$ 2.0 and 2.2), 
the solver is able to locate both the center and
radius of the offset horizons to an accuracy of better than 
0.06\% using general expansions to any order, $L=0$ to $L=6$.
We note that the iteration count (and hence cpu time) can
increase by factors of between 2 to 10, depending on the total
number of parameters,
as compared to the cases in which the throat center coordinates 
are not allowed to vary.

\subsubsection{Black Hole plus Brill Wave}
\label{sec:dbhinit}

The Misner initial data family just discussed provides examples of 
both single perturbed horizons and two separate but spherical
surfaces with offset centers for testing.  However, 
black hole horizons in highly dynamic spacetimes can be extremely 
distorted geometrically, 
and the horizon finder must be able to locate these as 
well.  The black hole plus Brill wave initial data set is yet another 
solution that has been studied extensively in axisymmetry, 
and thus provides a useful testbed for highly distorted holes in  
three dimensions.  This data describes the 
superposition of a black hole and a ``doughnut'' shaped Brill wave
surrounding the hole.  In spherical 
coordinates, the 3-metric takes the form
\begin{equation}
d\ell^2 = \psi^4 \left( e^{2q} \left( dr^2+r^2 d\theta^2\right) + r^2
\sin^2 \theta d\phi^2 \right),
\end{equation}
where $q$ and $\psi$ are functions of $r$ and $\theta$ only. The
function $q$ is specified analytically as free data,
and the Hamiltonian constraint
is solved for the conformal factor $\psi$. The initial extrinsic
curvature vanishes due to time symmetry.

This data set has been studied with 2D, axisymmetric
codes\cite{Bernstein93a,Bernstein94a} using a
logarithmic radial coordinate $\eta = \ln(2r/m_0)$, where $m_0$ is
a scale parameter. In this coordinate system ($\eta$, $\theta$, $\phi$),
the form of the $q$-function is written as
\begin{equation}
q = a \sin^n\theta g(\eta),
\end{equation}
where we set $n=2$, and
\begin{equation}
g(\eta) = \exp \left[ - \left(\frac{\eta+\eta_0}{\sigma}\right)^2
\right] + \exp \left[ - \left(\frac{\eta-\eta_0}{\sigma}\right)^2
\right].
\end{equation}
We solve the Hamiltonian constraint for the conformal factor
in our 2D code and then interpolate these solutions onto a 
$64^3$ Cartesian grid with $\Delta x = 0.15$ to
generate 3D data sets. The 3D horizon finder is tested against
an independent solver developed for 2D 
calculations~\cite{Bernstein93a}.

In Fig.~\ref{fig:dist_init} we show the 
coordinate location of the apparent horizon for various parameters of 
the $q$-function.  The sequence shown corresponds to different values 
of the Brill wave amplitude $a$, for fixed ``shape'' parameters having 
the values $(\eta_{0}, \sigma, n) = (0,1,2)$.  Results from both 
the 2D and 3D calculations are shown.  In the 3D case, we allowed 
searches up through $L = 4$, but for this case we include only the
axisymmetric terms. In all cases shown, the finder was able to 
locate the correct surface to within a third of a grid
zone in just 6 iterations.

Fig.~\ref{fig:dist_conv} is a resolution convergence study of the
case $a=-0.75$ in which the cell size is varied from
$\Delta x =$ 0.6, 0.3 and 0.15 with $16^3$, $32^3$ and $64^3$
grids respectively. The solver clearly converges to the correct surface
quadratically with cell size.

For negative values of the Brill wave amplitude parameter $a$, the 
horizon is found off the throat, but for positive values below certain 
limits (depending on the shape parameters) the horizon is located on 
the throat at $r=m_0/2$. We are also able to locate the AH at the same
level of accuracy for these cases.

Although the surfaces shown in Fig.~\ref{fig:dist_init}
appear to be almost spherical in coordinate space,
the presence of strong 
gravitational waves can severely distort the horizons geometrically,
much more so than the Misner data solutions described above.
In Fig. \ref{fig:embed1} we show the (axisymmetric) geometric embedding of 
the two cases $a=\pm 1$, 
using the method described in Ref. \cite{Anninos93a}.  
For all negative values of the amplitude parameter, the horizon is 
oblate. The $a = +1$ case is highly
distorted geometrically into a prolate shape, with a ratio of polar to 
equatorial circumference of $C_{p}/C_{e} = 4.28$.

\subsubsection{Kerr Black Hole}
\label{sec:kerr}

The calculations presented so far have tested the ability 
of the solver to find single or multiple horizons of
spherical and highly distorted black holes in 3D Cartesian 
coordinates.  However, the data in all these cases are time symmetric.
The Kerr initial data set describing a rotating black hole has 
nontrivial extrinsic curvature, and thus provides another important
testbed with a known analytic solution.
The 3-metric for this spacetime in Boyer-Lindquist 
coordinates is given by
\begin{equation}
d\ell^2 = \frac{\rho^2}{\Delta} dr^2 + \rho^2 d\theta^2 +
\frac{\left(\left(r^2+a^2\right)^2 - \Delta a^2
\sin^2\theta\right) \sin^2\theta}{\rho^2} d\phi^2,
\end{equation}
where
\begin{eqnarray}
\rho^2 &=& r^2 + a^2 \cos^2\theta , \\
\Delta &=& r^2 - 2 M r + a^2.
\end{eqnarray}
In these coordinates, the non-vanishing components of the 
extrinsic curvature are
\begin{eqnarray}
\hat{K}_{r\phi} &=& a M \left[ 2 r^2 \left( r^2 + a^2 \right) + \rho^2
\left( r^2 - a^2 \right) \right] \sin^2 \theta/\left(r \rho^4\right) , \\
\hat{K}_{\theta\phi} &=& -2 a^3 M r \sqrt{\Delta} \cos\theta
\sin^3\theta \rho^{-4} ,
\end{eqnarray}
where $\hat{K}_{ij}=\psi^2_0 K_{ij}$, and 
\begin{equation}
\psi^4_0 = \frac{\left(r^2+a^2\right)^2 - \Delta a^2
\sin^2\theta}{\rho^2}.
\end{equation}

To construct this data in 3D, we first transform to an isotropic
radial coordinate through the transformation
\begin{equation}
r = \bar{r} \left(1+\frac{M+a}{2\bar{r}}\right)
\left(1+\frac{M-a}{2\bar{r}}\right),
\end{equation}
as described in~\cite{Brandt94a}. In this coordinate system, the
coordinate singularities at $r=M \pm \sqrt{M^2-a^2}$ disappear. We
then convert the metric and extrinsic curvature from isotropic
coordinates to Cartesian coordinates.  The apparent horizon for the
Kerr data is a coordinate sphere, located at
\begin{equation}
\bar{r}=\frac{\sqrt{M^2-a^2}}{2}.
\end{equation}

In Table~\ref{tab:kerr_init} we show results from runs with various
values of the rotation parameter $a/M$ and expansion order $L$.  
These runs were all done with general expansions, without restricting
to axisymmetry. For the tests performed here, $\Delta x$,
$\Delta y$ and $\Delta z$ were chosen to be approximately one tenth of
the analytic AH radius, and the number of cells ($32^3$) was kept
the same in all cases.
Only the $\ell=0$ contributions are shown.
As expected, the contributions from higher-$\ell$ terms are small,
ranging from $10^{-9}$ for $a/M = 0$ to $3\times 10^{-2}$ for
$a/M=0.9$. The reason that these terms, in the highly rotating
cases, are not as small
as in the Schwarzschild case is because the metric is not
conformally flat, so the interpolations within the finder are not
as accurate. Although we do not display embeddings of these horizons,
we note that rotating black hole horizons can be extremely oblate, and
for rotation parameters $a/M > 0.866$ a global embedding into
Euclidean space is not possible.  The horizon finder has no
difficulty in locating these surfaces to within a grid zone, 
although we note, again, the high
computational cost for large order multipole expansions as
indicated by the $a/M=0.3$ sequence of calculations.

\subsubsection{Transformed Schwarzschild Black Hole}

The above test cases, although treated in full 3D, have all been
axisymmetric. In this section we test our finder on data which does
not have any particular symmetry, and for which we can derive the
correct solution in order to gauge the accuracy of the solver.
To this end, we chose to find the apparent horizon on a
Schwarzschild initial slice using a coordinate system in which
the horizon surface does not appear to be axisymmetric.
The coordinate transformation we use is 
\begin{equation}
\bar{r} = r f\left(\theta,\phi\right) 
   =  r \left({1+\frac{1}{4} \sin^2\theta \left( \cos^2\phi -
      \sin^2\phi \right)}\right)^{-1/2}
\end{equation}
where $r$ is the radial isotropic coordinate of (10).  This
$f\left(\theta,\phi\right)$ has surface modes of 
$L = 2$.  The apparent horizon location is then defined by
$\bar{r}=\frac{M}{2} f\left(\theta,\phi\right)$.

The solver was allowed to search coefficients with terms up to $L=2$
and $L=4$. In all cases for the $L=2$ tests, the finder successfully
found the horizon to high accuracy. Each non-zero coefficient was
accurate to better than a tenth of one percent, and the largest value
found for a coefficient that was supposed to be zero was less than 2\%
of the smallest non-zero component. We note that better accuracy can
be achieved if one uses full knowledge of the analytic form of the
metric. When the finder was generalized
to allow searches with coefficients up to $L=4$, the horizon was again
successfully found and to comparable accuracy.  However with the large
number of search parameters, some care was necessary in choosing the
initial search direction of the Powell routine in order to avoid
getting trapped at local minima in $\Theta^2$ generated in the
discretized representation of the spacetime.  When the resolution of
spacetime grid is increased, such local minima will diappear/change in
location, in sharp contrast to the actual $\Theta^2=0$ point (the actual
location of the AH, which converges to a fixed location with
increased resolution.)  In Fig.~\ref{fig:trnsch_init}, we show the
apparent horizon found and the analytic solution in one quadrant of
each of the coordinate planes.  The numerical results shown in this
figure were obtained with a run in which coefficients up to $L=4$ were
allowed to vary.

\subsection{Evolved 3D Data Sets}
\label{subsec:evolved}

The tests described thus far have shown that the horizon finder can
locate horizons in a variety of distorted black hole spacetimes, but
these have all been initial data sets with a somewhat restricted
3--metric and extrinsic curvature, and with a large part of the data
prescribed analytically.  In general 3D black hole evolutions, all
metric and curvature components will be present, and the data
will be contaminated with numerical inaccuracies generated during
the course of evolution. The horizon
finder must work under these conditions for it to be a useful tool 
in the numerical construction of spacetimes.
In this section, we discuss results derived from the
implementation of the horizon solver into our 3D ``G'' 
code \cite{Anninos94c} that
solves the Einstein evolution equations in Cartesian
coordinates.

\subsubsection{Schwarzschild}
\label{sec:schwarzevol}

The tests discussed here were carried out using multipole expansions
up to and including the $L=4$ terms. However, in order to
save computational time, we restricted the search to 
axisymmetric surfaces. We have verified that a more general expansion
does not change the results significantly.

The results for a 3D Schwarzschild spacetime, evolved with both
geodesic and maximal slicings with zero shift, are shown in Figs.
\ref{fig:sch_ev_geo} and \ref{fig:sch_ev_max} respectively. 
A $64^3$ ($130^3$) grid  with $\Delta x = $ 0.15 (0.2) was used
for the geodesic (maximal) case with $\Delta t = 0.25 \Delta x$.
Only the $\ell=0$
contribution to the surface is plotted. The other parameters remain small
($<10^{-3}$) during the entire evolutions, as expected.
In both cases, the surface locations are compared against the
corresponding results from 1D calculations.
The two results agree to a small fraction of a grid zone
throughout the evolution. 
The late time deviation in the maximal run is attributed to
errors in the 3D spacetime evolution, which becomes
inaccurate for $t > 20M$, although we note that the difference
at the end of the run is still only about a grid zone.

In both the geodesic and maximal cases, the AH finder accounted
for a large portion of the total cpu time.
For the geodesic case the code ran 160 timesteps,
invoking the finder once every $0.3M$ in time (11 times total).
In this case, the finder
constituted approximately 90\% of the cpu time. For the maximal
slicing case the code was run for 1100 timesteps, calling the finder
every $2M$ (14 times total).
In this case, 40\% of the cpu time was spent in finding the horizon.
The horizon finder with a $L=4$ axisymmetric multipole expansion 
is thus approximately 50--100 times slower than a single update
cycle in the hyperbolic evolution (although we note
that when the elliptic maximal
slicing equation is solved, the relative performance in the
solver improves significantly). A factor of roughly 10 can be
gained by reducing the expansion order to $L=0$.

\subsubsection{Misner 2BH Collision}
\label{sec:misnerevol}

A more difficult evolution scenario is to capture the horizon surfaces
as two black holes collide and merge. Our ``G'' code is used
to evolve the two black hole Misner data set, described 
in \S \ref{sec:misnerinit}, for a time sufficiently long that we can
test the AH finder on this dynamic spacetime.

To evolve the initial data, we use a zero shift vector and an
algebraic lapse of the form
\begin{equation}
\alpha = \alpha_o \left(1 + \log \hat\gamma \right) ,
\label{alglapse}
\end{equation}
where $\hat\gamma$ is the conformal 3--metric determinant and
$\alpha_o$ is the lapse on the initial time slice, which we
take to be the Cadez \cite{Cadez71} lapse
\begin{equation}
\alpha_o = \frac{1}{\psi} \left[1 + \sum_{n=1}^{\infty} (-1)^n
         \frac{1}{\sinh n\mu} 
         \left(\frac{1}{^+r_n} + \frac{1}{^-r_n}\right)\right] .
\label{cadezlapse}
\end{equation}
The solution (\ref{cadezlapse}) solves the maximal equation in
time symmetry with $\alpha = 0$ as a boundary condition on the
throats. Algebraic lapses of the form (\ref{alglapse}) are
singularity avoiding and produce evolutions similar to
maximally sliced spacetimes \cite{Anninos94c}. 

The calculation is run on a $64^3$ grid with $\Delta x = 0.1$ for a
time of $t=10M$ using an expansion to order $L=4$. 
As our interest is  in testing the ability of the
solver to locate the AH before and after the surface merger event,
it suffices to evolve a data set
with a low value for the Misner parameter, which is computationally less
expensive in evolution. Here we show
results for the $\mu=1.5$ case which has, as initial data,
two coalesced black holes with a common event horizon,
but two {\it distinct} trapped surfaces at the two throats. A common
apparent horizon encircling both throats forms at time
$t\sim 1.6M$.

Figure \ref{fig:misner_ev} plots the surfaces found at each time the
finder is called ($t=$ 0, 2.5, 5, 7.5 and 10$M$, where $M$ is the ADM
mass, and with the surfaces increasing in radius at the later times).
At $t=0$, the solver correctly finds the throat as the surface,
centered off the origin.  The finder can subsequently be prevented
from locking onto the throat (which remains a trapped surface
throughout the evolution) by restricting the range of radii over which
the function $F(\theta,\phi,r)$ is evaluated, and by resetting the
center of the surface.  Unfortunately, there is
no analytic result to compare against the computed location of the AH
at the later times.  Instead, we overlay the 3D computed AH surfaces
with the results from our 2D axisymmetric code evolving the same
initial data.  The surfaces determined with these two different
methods (3D minimization vs. ODE integration) applied to the two
different constructions of the spacetime (3D Cartesian evolution
vs. 2D axisymmetric evolution) basically agree with one another.
Nevertheless, we note that in the figure the two sets of surfaces are
not exactly the same in coordinate space, as they should {\it not }
be, since we are using different lapse and shift functions in the two
cases.  However, the shift vector is of magnitude $|\beta| \sim
10^{-3}$ in the region of the AH, so we do not expect this effect to
exceed the grid spacing scale over the time interval of the
calculation. Also, the algebraic lapse used in the 3D code is
typically larger than the maximal slicing lapse in the 2D code by a
fractional difference $\Delta \alpha \sim 0.02 $ near the AH.
This should also have a small effect on the coordinate position of the
AH except at late times. In fact, the maximum differences in the AH
location between the 2D and 3D calculations is only slightly more than
a grid cell.  The differences may also be attributed to inaccuracies
in the evolution; as born out in computing the mass of the
AH at different times.  At $t=2.5M$, after a common surface forms to
surround both throats, we find $M_{AH} = 2.28$, compared to the 2D
result of 2.36. However, by $t=10M$, the error in the mass of the
computed surface increases to roughly $18\%$ as the evolution becomes
less accurate. These errors are attributable to numerical inaccuracies
in the evolution and not the AH finder routine, in particular, they
are unrelated to the multipole order used, since the surface
becomes more spherical in time.

\subsubsection{Black Hole plus Brill Wave}
\label{subsubsec:ev_brill}

One final test of the solver in real time evolutions is a
Schwarzschild black hole plus a Brill wave.
The initial data is the same as that discussed in \S
\ref{sec:dbhinit} with shape parameters ($\eta_0$, $\sigma$, $n$) =
(0, 1, 2), scale parameter $m_0=2$ and amplitude $a=-0.5$. For this set
of parameters, the ADM mass of the spacetime is $M=1.77m_0 = 3.54$. The
Hamiltonian constraint is solved for the conformal factor in our 2D
axisymmetric code, then interpolated onto a $66^3$ Cartesian grid with
$\Delta x = 0.2 = 0.056M$. The data is evolved with a timestep $\Delta
t = 0.25 \Delta x = 0.014M$, calling the horizon finder once every
$m_0=0.56M$ intervals of time. The shift vector is set to zero and the
lapse function is computed from the maximal slicing condition.

Figure \ref{fig:ev_brill_surf} shows the horizon shapes and locations
in the 3D calculation at various times with intervals of $2m_0=1.15M$,
starting at $t=0$.  The multipole order used in this calculation is
$L=4$, with $m=0$ to reduce the computational time. The higher order
expansion is needed here to describe the oblate shape of the horizons.
For the most part, the solver required only about five iterations to
converge when the previous solution is given as the initial guess to
the finder.  Also shown in Fig. \ref{fig:ev_brill_surf} are the
corresponding surfaces found in our 2D axisymmetric code.  Once more
we note that the slicing and shift conditions differ in the two cases,
so we do not expect the surfaces to coincide precisely.  However as
the differences between the lapse functions and shift vectors are
small ($|\beta| \sim 10^{-3}$ and $\Delta \alpha \sim 0.07$ in regions
near the horizon except near the $z$--axis), the solutions do agree
nicely.  In and around the $x-y$ plane, the solutions match to within
half a grid cell.  Greater differences, however, are found along the
$z$--axis, where the two surfaces are displaced by a maximum of
roughly two grid cells. This is attributed in part to a bigger $\Delta
\alpha$ ($\sim 0.1$ ) near the $z$--axis, which results from the
imposed asymmetry in the lapse function due to the nearness of the
outer boundaries, where we enforce the spherical Schwarzschild lapse
as a boundary condition in the maximal equation used in the 2D
simulation.

Again, a more geometrical comparison or test of the solver is the mass of the
surface found. The horizon mass is defined as 
$M_{AH} = \sqrt{A_{AH}/16\pi}$, where $A_{AH}$ is the area of the
surface. Figure \ref{fig:ev_brill_mass} plots the AH mass as a function
of time for the 2D and the $L=4$, 3D evolutions. In both cases, the mass
increases at first, as the gravitational waves fall into the black hole,
reaching $M_{AH} \sim 0.997 M$ at $t\sim 6M$.
The masses in the two cases differ by only 0.1\% at $t\sim 6M$.
For comparison we also plot $M_{AH}$ for the surface found using a lower
order $L=2$ multipole expansion. At early times, when the horizon
is most distorted, the $L=2$ expansion is clearly not adequate to
resolve the horizon shape, as evidenced by the AH mass which exceeds
the ADM mass by about 1\%. However, as the black hole settles down into
a quasi--static state, the surface becomes more spherical and the
$L=2$ solution approaches both the 2D and the $L=4$, 3D results,
differing from the 2D result by about 0.35\% at $t\sim 6M$.

\section{Conclusions}
\label{sec:conclusion}

We have developed a promising general 3D method of finding the AH in a
numerically constructed spacetime based on a minimization procedure. 
In this paper we have applied the method and demonstrated that it works
for spacetime data which are not time symmetric, and data sets which are
obtained in actual numerical evolutions.

The major advantage of this method is that the minimization procedure is
much better understood than the corresponding problem of solving the
elliptic equation, and well--tested routines are available for
solutions.  The major drawback of the method is that, for AH surfaces
which are not smooth, and which deviate significantly from sphericity
in coordinate space, 
the higher dimensional minimization procedure
can be computationally expensive.  
With our present implementation
of the finder using Powell's routine, we are limited to searching for
AH surfaces at every 50, or so, evolution cycles, instead of continually
monitoring the AH throughout the numerical evolution, as would be
our final goal.  In future implementations of the method, we anticipate
developing a more sophisticated minimization routine 
using derivative information to speed up the procedure, as well as
other means to improve the robustness and efficiency of the code.
Our code, which is optimized for both the C90 and massively
parallel CM5 machines, together with documentation, 
will be made available on our
servers http://jean-luc.ncsa.uiuc.edu and
http://wugrav.wustl.edu. 

\section*{Acknowledgements} We gratefully acknowledge the assistance 
of Andrey Odintsov and David Rosnick, who helped in testing and coding 
various components of the code described in this paper.  We also
thank Thomas Baumgarte for discussions and John Shalf for help in
visualization of the horizon surfaces presented in this paper.  This 
research is supported by the National Center for Supercomputing
Applications, the Pittsburgh Supercomputing 
Center, and NSF grants Nos. PHY94-04788, PHY94-07882, PHY96-00507 and 
PHY/ASC93-18152 (arpa supplemented).  WMS would also like to thank the 
Institute of Mathematical Sciences of The Chinese University of Hong Kong
for hospitality during his visit.



\begin{figure}
\caption{
The coordinate location of the apparent horizon in
the $x$--$z$ plane for Schwarzschild initial data with various
masses. The computed surfaces are indistinguishable from the
analytic solutions, with deviations of order 0.04\%.
}
\label{fig:sch_init}
\end{figure}

\begin{figure}
\caption{
The coordinate location of the apparent horizon for the
$\mu=1.2$ Misner initial data. The solid line is the apparent horizon
computed from our 2D code. The various broken lines are the
surfaces obtained from the different multipole order expansions.
The surface obtained from the $L=6$ expansion is indistinguishable
from the 2D result in this plot. We note that terms up
to $L=6$ must be included to find the AH accurately.
}
\label{fig:mis_init}
\end{figure}

\begin{figure}
\caption{
The coordinate location of the apparent horizon for 2D and 3D
Brill wave plus black hole initial data with various values of the
Brill wave amplitude. In all runs, the Brill wave is centered at
$\eta_0=0$ and has a width of $\sigma=1$. Note that there is good
agreement even though the initial data is not analytic and actually 
contains highly geometrically distorted black hole horizons.
}
\label{fig:dist_init}
\end{figure}

\begin{figure}
\caption{
A resolution study of the Brill wave plus black hole spacetime
for the case ($a, \eta_0, \sigma$) = ($-0.75, 0, 1$). The
surface converges quadratically with grid spacing
to the correct location, as
represented by an independent 2D calculation shown by the
solid line.
}
\label{fig:dist_conv}
\end{figure}

\begin{figure}
\caption{
The geometric embedding diagrams of the apparent
horizons for the Brill wave plus black hole initial data with
wave amplitudes $a=\pm 1$. The $a=-1$ horizon is quite oblate,
and deviates from sphericity even in its coordinate location.
For the $a=1$ case, the
horizon is actually located on the throat, which is a coordinate
sphere. However, the metric functions are highly nonspherical, 
leading to this very prolate geometry of the horizon surface.
}
\label{fig:embed1}
\end{figure}

\begin{figure}
\caption{
Apparent horizon location in each of the coordinate planes for the
transformed Schwarzschild initial data. The numerical data are
represented by solid lines, and the analytic data by symbols. The
surface is pure $L=2$, although the finder was allowed to search
through the $L=4$ coefficients as well. 
The computed coefficients are found to roughly 0.1\% accuracy.
}
\label{fig:trnsch_init}
\end{figure}

\begin{figure}
\caption{
The apparent horizon location for 1D and 3D
Schwarzschild evolutions using geodesic slicing. The 1D data were
obtained using 128 grid points and resolution $\Delta r =
0.0375M$, where $M$ is the mass of the black hole. 
The 3D data were obtained using $64^3$ grid points and
resolution $\Delta x = 0.075M$. Only the $\ell=0$ contribution to the
3D apparent horizon is plotted but the other terms in the
$L=4$ series are negligible, as
discussed in the text.}
\label{fig:sch_ev_geo}
\end{figure}

\begin{figure}
\caption{
The apparent horizon location for 1D and 3D
Schwarzschild evolutions using maximal slicing. The 1D data were
obtained using 130 grid points and resolution $\Delta r = 0.1M$, the
3D data using $130^3$ grid zones and resolution $\Delta
x = 0.1M$. We note that the agreement is within a grid zone even 
to the end of the calculation, when the 3D evolution becomes
inaccurate.}
\label{fig:sch_ev_max}
\end{figure}

\begin{figure}
\caption{
Apparent horizon locations in the 3D evolution of the
$\mu=1.5$ Misner two black hole data. The surfaces are plotted
at times $t=$ 0, 2.5, 5, 7.5 and 10$M$, where $M$ is the ADM
mass of the spacetime, and the surfaces increase in radius with time.
The evolution is performed
using a $64^3$ grid with resolution $\Delta x = 0.1$ and
a multipole expansion of order $L=4$.
Also shown are the corresponding horizons found in the
2D axisymmetric evolutions. The surfaces from the
two calculations agree nicely, although we note that
an exact correspondence is not expected due to different shift
and lapse functions.
}
\label{fig:misner_ev}
\end{figure}

\begin{figure}
\caption{
Coordinate location of the AH found in the 2D and 3D evolutions of the
Brill wave plus black hole spacetime. The surfaces are shown starting at
$t=0$ with
time intervals of $2m_0=1.12M$, where $M=3.54$ is the ADM mass of the
spacetime. Although we do not expect identical results due to
different kinematic conditions in the two cases, the 
surfaces differ at most by little more than two grid cells.
The evolution is performed on a $66^3$ grid with $\Delta x = 0.2$.
}
\label{fig:ev_brill_surf}
\end{figure}

\begin{figure}
\caption{
Comparison of the apparent horizon masses computed from the 2D and 3D
evolutions of the Brill wave plus black hole spacetime. The AH
mass increases initially as the Brill wave falls into the black hole.
By $t\sim 6M$, the mass approaches $0.997M$ and the 2D and 
$L=4$, 3D results differ by just 0.1\% at this time.
We also show the corresponding masses computed from the surfaces found
with an $L=2$ multipole expansion. The low order expansion is clearly
not adequate in resolving the surface at early times when
the horizon is most distorted.
}
\label{fig:ev_brill_mass}
\end{figure}


\begin{table}
\begin{tabular}{cccc}
$L$ & iterations & cpu time & $\sum\Theta^2$ \\ \hline
0 & 2  & 1.2 min & $8.3\times 10^{-2}$ \\
2 & 5  & 2.2 min & $8.0\times 10^{-3}$ \\
4 & 9  & 4.6 min & $7.9\times 10^{-4}$ \\
6 & 16 & 12  min & $1.0\times 10^{-4}$ \\
\end{tabular}
\caption{
The effect of the (axisymmetric) multipole 
order $L$ on finding the
correct horizon surface is demonstrated here for the
two black hole Misner data with $\mu=1.2$. The number of
iterations and cpu time required by the solver is tabulated
along with the expansion summed over all points on the surface.
The timings were performed on the 128-node partition of the CM5,
using a 3D grid of size $64^3$ and a $5\times 5$ mesh to cover
the 2D horizon surface in a single quadrant.
}
\label{tab:misner}
\end{table}

\begin{table}
\begin{tabular}{cccccccc}
$a/M$ & $L$ & $\Delta x$ & $r_{a}$ & $r_{n}$ & 
$\Delta r/\Delta x$ & iterations & cpu  \\ \hline
0   & 4 & 0.1   & 1.0   & 1.0002 & 0.003 & 30 & 2.4 hr  \\
0.3 & 4 & 0.1   & 0.954 & 1.0095 & 0.556 & 37 & 1.0 hr  \\
0.3 & 2 & 0.1   & 0.954 & 0.9562 & 0.022 & 9  & 6.7 min  \\
0.3 & 0 & 0.1   & 0.954 & 0.9542 & 0.002 & 2  & 0.5 min  \\
0.9 & 4 & 0.039 & 0.436 & 0.4748 & 0.997 & 27 & 0.5 hr  \\
\end{tabular}
\caption{
Performance measures of the horizon solver applied to the
Kerr black hole data with various spin parameters $a/M$.
Here $L$ is the maximum multipole order, $\Delta x$ is the
3D grid spacing, $r_{a}$ is the analytical position of
the horizon, $r_{n}$ is the horizon location found by the
solver, and $\Delta r/\Delta x$ is the difference between the analytic
and numerical locations normalized to the grid spacing.
We also show the number of iterations and cpu time required by
the solver running on a 32-node partition of the CM5, using a grid with
$32^3$ cells and a $5\times 5$ mesh for the surface.
}
\label{tab:kerr_init}
\end{table}

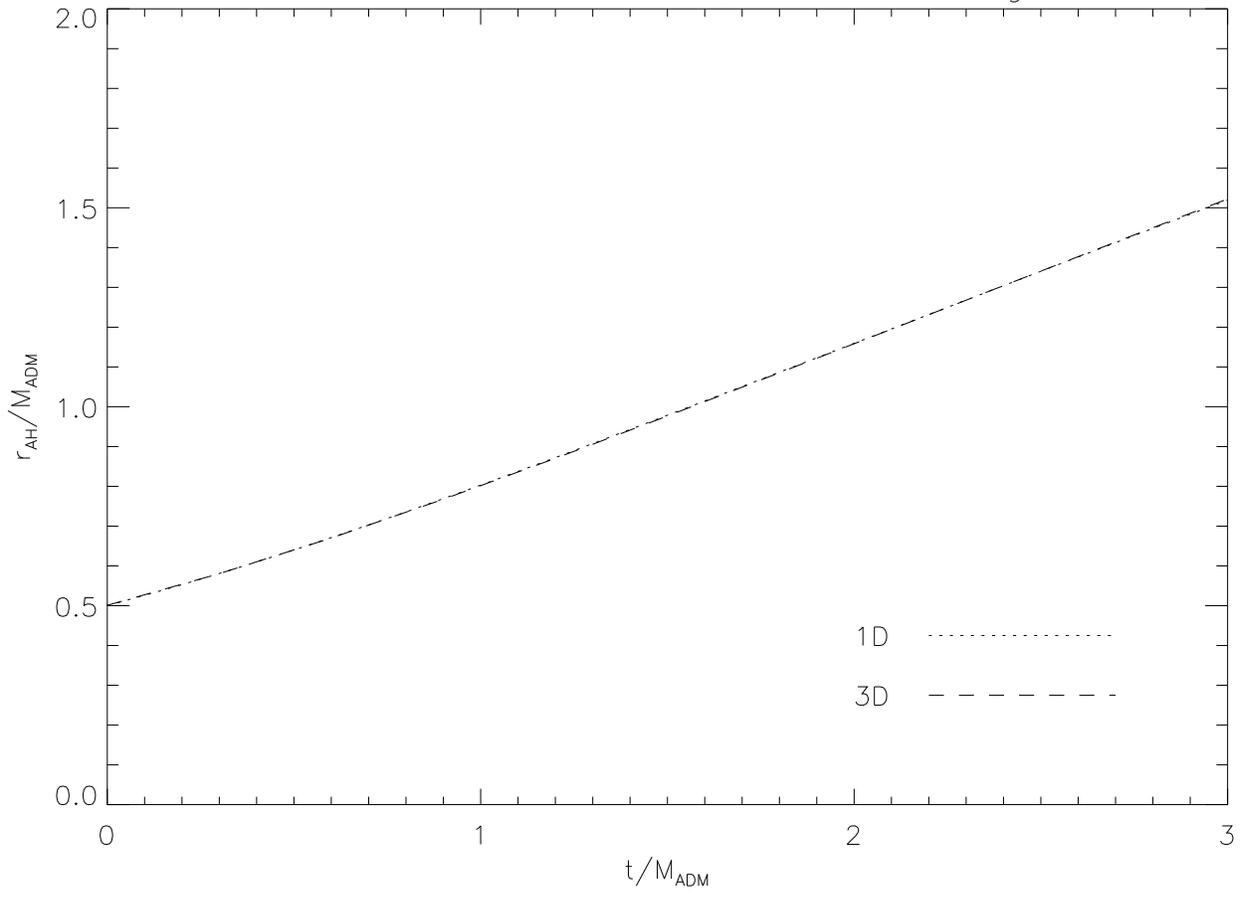
\end{document}